\documentclass[11pt]{article}

\usepackage{amsfonts,amssymb,chet,dsfont,graphicx,mathrsfs,pgfplots,tikz}

\usepackage{subcaption}
\usetikzlibrary{external}
\title{Probability Density Functions for\\CP-Violating Rephasing Invariants}

\author{Jean-Fran\c{c}ois Fortin\email{jean-francois.fortin@phy.ulaval.ca}, Nicolas Giasson\email{nicolas.giasson.1@ulaval.ca} and Luc Marleau\email{luc.marleau@phy.ulaval.ca}}

\affiliation{
D\'epartement de Physique, de G\'enie Physique et d'Optique,\\Universit\'e Laval, Qu\'ebec, QC G1V 0A6, Canada
}

\abstract{%
The implications of the anarchy principle on CP violation in the lepton sector are investigated.  A systematic method is introduced to compute the probability density functions for the CP-violating rephasing invariants of the PMNS matrix from the Haar measure relevant to the anarchy principle.  Contrary to the CKM matrix which is hierarchical, it is shown that the Haar measure, and hence the anarchy principle, are very likely to lead to the observed PMNS matrix.  Predictions on the CP-violating Dirac rephasing invariant $|j_D|$ and Majorana rephasing invariant $|j_1|$ are also obtained.  They correspond to $\langle|j_D|\rangle_\text{Haar}=\pi/105\approx0.030$ and $\langle|j_1|\rangle_\text{Haar}=1/(6\pi)\approx0.053$ respectively, in agreement with the experimental hint from T2K of $|j_D^\text{exp}|\approx0.032\pm0.005$ (or $\approx0.033\pm0.003$) for the normal (or inverted) hierarchy.
}

\date{January 2018} 

\begin{document}

\maketitle



\section{Introduction}\label{SIntro}

Although the Standard Model (SM) of particle physics is well established, it must be extended to account for neutrino oscillations \cite{pdg,Esteban:2016qun}.  With this extension comes the possibility of flavor violation in the lepton sector encoded in the Pontecorvo-Maki-Nakagawa-Sakata (PMNS) matrix, which is the equivalent of the Cabibbo-Kobayashi-Maskawa (CKM) matrix of the quark sector.  Experimental data show that, contrary to the quark sector where the CKM matrix is quite hierarchical, the PMNS matrix prefers near-maximal mixing.

One possible physical motivation for the peculiar pattern of the PMNS matrix is the anarchy principle \cite{Hall:1999sn,deGouvea:2003xe,Heeck:2012fw,deGouvea:2012ac,Bai:2012zn,Lu:2014cla,Babu:2016aro,Long:2017dru,Haba:2000be,Espinosa:2003qz,Fortin:2016zyf,Fortin:2017iiw}.  Under the anarchy principle, the light neutrino mass matrix is generated by one of the seesaw mechanisms while the fundamental high-energy Dirac and Majorana neutrino mass matrices are obtained randomly from appropriate Gaussian ensembles.  The implications of the anarchy principle for the light neutrino mass spectrum were studied analytically in \cite{Fortin:2016zyf,Fortin:2017iiw} where it was proved that the probability density function (pdf) factorizes in a pdf for the light neutrino masses and a pdf for the mixing angles and CP-violating phases of the PMNS matrix.  This factorization property results in a lack of correlation between the light neutrino masses on one side and the mixing angles and CP-violating phases on the other.  It was thus argued in \cite{Fortin:2016zyf,Fortin:2017iiw} that the anarchy principle cannot constrain the neutrino hierarchy (normal or inverted), only the preferred neutrino mass splitting $m_\text{med}^2-m_\text{min}^2>m_\text{max}^2-m_\text{med}^2$ or $m_\text{med}^2-m_\text{min}^2<m_\text{max}^2-m_\text{med}^2$.  Moreover, it was shown that the experimentally-favored seesaw mechanism corresponds to type I-III for which the preferred neutrino mass splitting matches the normal hierarchy.

Although the pdf for the light neutrino mass spectrum depends on the type of seesaw mechanism, it was shown mathematically in \cite{Fortin:2016zyf} that the pdf for the mixing matrix does not (this was already foreseen in \cite{Haba:2000be} based on physical arguments).  Indeed, in the general $N\times N$ case, the pdf for the $N\times N$ unitary matrix is simply the Haar measure of the unitary group $U(N)$ which, when focusing on the mode, prefers near-maximal mixing.  However, since the mode is not invariant under changes of variables \cite{Espinosa:2003qz}, an appropriate probability test comparing the Haar measure and the uniform measure was done in \cite{Fortin:2017iiw}.  It was shown there that the probability test cannot discriminate between the two hierarchies and that the Haar measure is approximately two times more likely to generate mixing angles in the allowed experimental regions ($1\sigma$) than the uniform measure.

At first sight, the (joint) pdf for the PMNS matrix, given by the Haar measure, does not seem to lead to meaningful constraints on the CP-violating Dirac and Majorana phases.  Indeed, from the Haar measure the pdf of any CP-violating phase is uniform.  However, both the CKM and the PMNS matrices can be modified by phase rotations of the appropriate fermion fields.  Since physical quantities must be invariant under change of basis, the appropriate quantities to study from the Haar measure are not the CP-violating phases but the rephasing invariants \cite{Jarlskog:1985ht,Greenberg:1985mr,Dunietz:1985uy}.  The Jarlskog invariant $j_D$ \cite{Jarlskog:1985ht} associated to the CP-violating Dirac phase of the CKM matrix was studied statistically in \cite{Gibbons:2008su,Dunkl:2009sn}.  The pdf for the Jarlskog invariant $j_D$ was computed analytically in \cite{Dunkl:2009sn} and it was shown there that the CKM matrix should not be considered a generic unitary matrix obtained randomly from the Haar measure.

Since an equivalent investigation has not been performed for the lepton sector, it is the goal of this paper to fill this gap.  As will be seen below, the statistical evidence of the Haar measure for the PMNS matrix is much stronger than for the CKM matrix, especially with the recent hint for a CP-violating Dirac phase $\delta$ around $270^\circ$, mostly coming from T2K \cite{Abe:2017uxa,Abe:2017vif}, that leads to the lepton sector Jarlskog invariant $j_D^\text{exp}\approx-0.032\pm0.005$ (or $\approx-0.033\pm0.003$) for the normal (or inverted) hierarchy, which is three orders of magnitude larger than the quark sector Jarlskog invariant.

This paper is organized as follows: Section \ref{SRInv} discusses rephasing invariants, mostly the quartic rephasing invariants related to CP violation coming from the Dirac phases (relevant to both the quark sector and the lepton sector) and the Majorana phases (possibly relevant only to the lepton sector).  The cases $N=2$ and $N=3$ are shown explicitly with both the Haar measure and the uniform measure.\footnote{Although the uniform measure is not physically motivated, it is seen as a na\"ive benchmark hypothesis to which the Haar measure can be compared to.}  In section \ref{SPDFs} some preliminary results are introduced to simplify the computations of the different moments of the CP-violating rephasing invariants and of the associated pdfs.  The pdfs for properly normalized CP-violating rephasing invariants in the cases $N=2$ and $N=3$ are then obtained with the help of the pdfs of the products of beta-distributed random variables.  Section \ref{SDis} presents a discussion of the results including a comparison with numerical results.  In the physical neutrino case with $N=3$, it is shown that there is strong statistical evidence that the PMNS matrix is a generic unitary matrix drawn at random from the statistical ensemble associated to the Haar measure, contrary to the CKM matrix.  Indeed, the average of the Jarlskog invariant is $\langle|j_D|\rangle_\text{Haar}=\pi/105\approx0.030$ for the Haar measure, which agrees well with $|j_D^\text{exp}|\approx0.032\pm0.005$ (or $\approx0.033\pm0.003$) for the normal (or inverted) hierarchy.  Before concluding in section \ref{SConclusion}, some implications for CP violation are also presented.


\section{Rephasing invariants}\label{SRInv}

In the mass-eigenstate basis of the quark and lepton sectors, flavor violation is parametrized by the CKM and the PMNS (unitary) matrices.  There is however still some freedom associated with phase rotations of the fermion fields.  This freedom leads to a redefinition, more precisely a rephasing, of the CKM and the PMNS matrices.  Since physical quantities are basis-independent, they must be invariant under this rephasing.  The rephasing invariants \cite{Jarlskog:1985ht,Greenberg:1985mr,Dunietz:1985uy} of the CKM and the PMNS matrices are either CP-conserving or CP-violating.  The focus here is on the $N(N-1)/2$ CP-violating rephasing invariants of the PMNS matrix associated with the $N(N-1)/2$ CP-violating phases of the $N\times N$ case.

The exact form of the CP-violating rephasing invariants of the PMNS matrix $U$ in terms of the mixing angles and phases is explicitly known for general $N$.  They are given by quartic invariants of the PMNS matrix elements \cite{Jenkins:2007ip},
\eqna{
j_{D,(i-1,j-1)}&=\text{Im}(U_{11}U_{ij}U_{1j}^*U_{i1}^*),\qquad2\leq i\leq j\leq N-1,\\
j_{i-1}&=\text{Im}(U_{1i}U_{1i}U_{11}^*U_{11}^*),\qquad2\leq i\leq N,
}[EqnCPoddInv]
and can be expressed in terms of the mixing angles and phases with a suitable parametrization for the PMNS matrix.  Technically, when the CP-conserving quadratic and quartic rephasing invariants are taken into account, only discrete information about the signs of the sine of the CP-violating phases is encoded in the CP-violating invariants \eqref{EqnCPoddInv} \cite{Jenkins:2007ip}.  As will be seen below, the information about these signs is unfortunately lost in the statistical analysis presented here, due to the invariance of the pdfs under parity $j\to-j$ for all CP-violating rephasing invariants \eqref{EqnCPoddInv}.  However, for the physical case of neutrino physics with $N=3$, the pdfs for the absolute values of the CP-violating rephasing invariants will nevertheless permit to predict the value of one of the CP-violating Majorana phases (up to the aforementioned sign) which is still unknown experimentally.

\subsection{\texorpdfstring{$N=2$}{N=2}}\label{SsRInv2}

One possible parametrization of $2\times2$ unitary matrices (with unphysical phases removed) is given by
\eqn{U=\left(\begin{array}{cc}\cos(\theta)&-\sin(\theta)e^{i\gamma}\\\sin(\theta)e^{-i\gamma}&\cos(\theta)\end{array}\right),}[EqnU2]
where $\theta\in[0,\pi/2]$ and $\gamma\in[0,2\pi]$.  Since there is only one phase in \eqref{EqnU2}, there exists only one CP-violating rephasing invariant for $N=2$, as expected from \eqref{EqnCPoddInv}.  It can be expressed as
\eqn{j=\text{Im}(U_{12}U_{12}U_{11}^*U_{11}^*)=2\sin^2(\theta)\cos^2(\theta)\sin(\gamma)\cos(\gamma),}[Eqnj2]
where the last equality is obtained from the parametrization \eqref{EqnU2}.  Here $j$ is defined on $[-1/4,1/4]$.  The (normalized) measures that will be studied later are the Haar measure and the uniform measure,
\eqn{d\mu_\text{Haar}=\frac{1}{\pi}\sin(\theta)\cos(\theta)d\theta d\gamma,\qquad\qquad d\mu_\text{Uniform}=\frac{1}{\pi^2}d\theta d\gamma.}[EqnMeasure2]
The statistical implications of these measures will be compared in the last section.  However, it is already clear due to parity that the odd moments of the rephasing invariant \eqref{Eqnj2} vanish for both measures \eqref{EqnMeasure2}.

\subsection{\texorpdfstring{$N=3$}{N=3}}\label{SsRInv3}

Although it is not necessary to use the PMNS parametrization \cite{pdg} for $3\times3$ unitary matrices, it is convenient to visualize the CP-violating Dirac phase $\delta$ and the two CP-violating Majorana phases $\alpha_{21}$ and $\alpha_{31}$.  Hence $N=3$ unitary matrices are parametrize such as
\eqna{
U&=\left(\begin{array}{ccc}1&0&0\\0&\cos(\theta_{23})&\sin(\theta_{23})\\0&-\sin(\theta_{23})&\cos(\theta_{23})\end{array}\right)\left(\begin{array}{ccc}\cos(\theta_{13})&0&\sin(\theta_{13})e^{-i\delta}\\0&1&0\\-\sin(\theta_{13})e^{i\delta}&0&\cos(\theta_{13})\end{array}\right)\\
&\phantom{=}\hspace{0.5cm}\times\left(\begin{array}{ccc}\cos(\theta_{12})&\sin(\theta_{12})&0\\-\sin(\theta_{12})&\cos(\theta_{12})&0\\0&0&1\end{array}\right)\left(\begin{array}{ccc}1&0&0\\0&e^{i\alpha_{21}/2}&0\\0&0&e^{i\alpha_{31}/2}\end{array}\right),
}[EqnU3]
where the mixing angles $\theta_{12}$, $\theta_{13}$ and $\theta_{23}$ are defined on $[0,\pi/2]$ and the phases $\delta$, $\alpha_{21}$ and $\alpha_{31}$ are defined on $[0,2\pi]$.  All unphysical phases are neglected in \eqref{EqnU3}.  The three CP-violating rephasing invariants associated to the three phases are
\eqna{
j_D&=\text{Im}(U_{11}U_{22}U_{12}^*U_{21}^*)=\sin(\theta_{12})\cos(\theta_{12})\sin(\theta_{13})\cos^2(\theta_{13})\sin(\theta_{23})\cos(\theta_{23})\sin(\delta),\\
j_1&=\text{Im}(U_{12}U_{12}U_{11}^*U_{11}^*)=\sin^2(\theta_{12})\cos^2(\theta_{12})\cos^4(\theta_{13})\sin(\alpha_{21}),\\
j_2&=\text{Im}(U_{13}U_{13}U_{11}^*U_{11}^*)=\cos^2(\theta_{12})\sin^2(\theta_{13})\cos^2(\theta_{13})\sin(\alpha_{31}-2\delta),\\
}[Eqnj3]
where the last equalities are obtained from the PMNS parametrization \eqref{EqnU3}.  Here the rephasing invariants are defined such that $j_D\in[-1/(6\sqrt{3}),1/(6\sqrt{3})]$, $j_1\in[-1/4,1/4]$ and $j_2\in[-1/4,1/4]$ respectively.  The (normalized) measures of interest are the Haar measure and uniform measure.  They are given by
\eqna{
d\mu_\text{Haar}&=\frac{2}{\pi^3}\sin(\theta_{12})\cos(\theta_{12})\sin(\theta_{13})\cos^3(\theta_{13})\sin(\theta_{23})\cos(\theta_{23})d\theta_{12}d\theta_{13}d\theta_{23}d\delta d\alpha_{21}d\alpha_{31},\\
d\mu_\text{Uniform}&=\frac{1}{\pi^6}d\theta_{12}d\theta_{13}d\theta_{23}d\delta d\alpha_{21}d\alpha_{31}.
}[EqnMeasure3]
Again, it is already possible to verify that the odd moments of the rephasing invariants \eqref{Eqnj3} vanish for both measures \eqref{EqnMeasure3} due to parity.  As mentioned before, the Haar measure is well motivated by physical arguments.  The uniform measure will be used to perform a statistical comparison in the physical neutrino case.


\section{Probability Density Functions}\label{SPDFs}

The main results of this paper are included in this section.  After introducing some preliminary results, the non-vanishing moments are computed for the rephasing invariants.  They are then shown to be the same than the moments of products of random variables with well-known pdfs.  The relevant rephasing invariant pdfs are finally obtained analytically from the products of random variables.

\subsection{Preliminary Results}\label{SsPrel}

To simplify the proofs of the CP-violating rephasing invariant pdfs, it is of interest to introduce some preliminary results.

First, with the change of variables $x=\sin^2(\theta)$, the following integrals are given by
\eqna{
\int_0^{\pi/2}\sin^i(\theta)\cos^j(\theta)d\theta&=\frac{1}{2}\int_0^1x^{(i+1)/2-1}(1-x)^{(j+1)/2-1}dx=\frac{\Gamma[(i+1)/2]\Gamma[(j+1)/2]}{2\Gamma[(i+j+2)/2]},\\
\int_0^{2\pi}\sin^{2k}(\theta)\cos^{2\ell}(\theta)d\theta&=2\xi_k\xi_{k+\ell}\int_0^1x^{(k+1/2)-1}(1-x)^{(\ell+1/2)-1}dx=\frac{2\xi_k\xi_{k+\ell}\Gamma(k+1/2)\Gamma(\ell+1/2)}{\Gamma(k+\ell+1)}.
}[EqnInt]
where $\text{Re}(i)>-1$, $\text{Re}(j)>-1$, $\text{Re}(k)>-1/2$, $\text{Re}(\ell)>-1/2$ and $\xi_k=[1+(-1)^{2k}]/2$.  These integrals can be used to compute the non-vanishing moments of the rephasing invariants.

Then, to write the rephasing invariants in terms of products of random variables, it is necessary to introduce the pdf for a beta-distributed random variable $X$ defined on $[0,1]$, which is
\eqn{P_{(\alpha,\beta)}(x)=\frac{\Gamma(\alpha+\beta)}{\Gamma(\alpha)\Gamma(\beta)}x^{\alpha-1}(1-x)^{\beta-1},}
where $\text{Re}(\alpha)>0$ and $\text{Re}(\beta)>0$.  Hence, the moments of $X$ are
\eqn{\langle x^n\rangle_{(\alpha,\beta)}\equiv\int_0^1x^nP_{(\alpha,\beta)}(x)dx=\frac{(\alpha)_n}{(\alpha+\beta)_n},}[EqnRV]
where $n\in\mathbb{N}$ and $(\alpha)_n\equiv\Gamma(\alpha+n)/\Gamma(\alpha)$ is the Pochhammer symbol.  The duplication formula for the Pochhammer symbol,
\eqn{(\alpha)_{mn}=m^{mn}\prod_{i=0}^{m-1}\left(\frac{\alpha+i}{m}\right)_n,}[EqnDup]
with $m,n\in\mathbb{N}$, is also necessary to express \eqref{EqnInt} in terms of \eqref{EqnRV}.

Finally, for two independent random variables $X$ and $Y$ defined on $[0,1]$ with pdfs $P_X$ and $P_Y$ respectively, the pdf for the product $XY$ is simply given by
\eqn{P_{XY}(x)=\int_x^1P_X(t)P_Y(x/t)\frac{1}{t}dt.}
Hence the pdf for the product of two random variables with pdfs $P_{(\alpha_1,\beta_1)}$ and $P_{(\alpha_2,\beta_2)}$ respectively is
\eqna{
P_{(\alpha_1,\beta_1;\alpha_2,\beta_2)}(x)&=\int_x^1P_{(\alpha_1,\beta_1)}(t)P_{(\alpha_2,\beta_2)}(x/t)\frac{1}{t}dt\\
&=\frac{\Gamma(\alpha_1+\beta_1)\Gamma(\alpha_2+\beta_2)}{\Gamma(\alpha_1)\Gamma(\alpha_2)\Gamma(\beta_1)\Gamma(\beta_2)}x^{\alpha_2-1}\int_x^1t^{\alpha_1-\alpha_2-\beta_2}(1-t)^{\beta_1-1}(t-x)^{\beta_2-1}dt\\
&=\frac{\Gamma(\alpha_1+\beta_1)\Gamma(\alpha_2+\beta_2)}{\Gamma(\alpha_1)\Gamma(\alpha_2)\Gamma(\beta_1+\beta_2)}x^{\alpha_2-1}(1-x)^{\beta_1+\beta_2-1}{}_2F_1(\beta_1,\beta_2-\alpha_1+\alpha_2;\beta_1+\beta_2;1-x),
}[EqnPDFab]
where ${}_2F_1(a,b;c;z)$ is the Gaussian hypergeometric function.  With the help of the Euler transformation of Gaussian hypergeometric functions ${}_2F_1(a,b;c;z)=(1-z)^{c-a-b}{}_2F_1(c-a,c-b;c;z)$, \eqref{EqnPDFab} is invariant under $(\alpha_1,\beta_1)\leftrightarrow(\alpha_2,\beta_2)$ as expected by symmetry.  The proof of \eqref{EqnPDFab} is a straightforward generalization of the proof given in Appendix B of \cite{Dunkl:2009sn}.  Moreover, the pdf for the product of two random variables with pdfs $P_{(\alpha_1,\beta_1;\alpha_2,\beta_2)}$ and $P_{(1/2,1)}$ respectively, such that $\beta_1+\beta_2=1$, is given by
\eqna{
P_{(\alpha_1,\beta_1;\alpha_2,\beta_2;1/2,1)}(x)&=\int_x^1P_{(\alpha_1,\beta_1;\alpha_2,\beta_2)}(t)P_{(1/2,1)}(x/t)\frac{1}{t}dt\\
&=\frac{\Gamma(\alpha_1+\beta_1)\Gamma(\alpha_2+\beta_2)}{2\Gamma(\alpha_1)\Gamma(\alpha_2)\sqrt{x}}\int_x^1t^{\alpha_2-3/2}{}_2F_1(\beta_1,\beta_2-\alpha_1+\alpha_2;1;1-t)dt\\
&=\frac{\sin(\pi\beta_1)\sin[\pi(\beta_2-\alpha_1+\alpha_2)]\Gamma(\alpha_1+\beta_1)\Gamma(\alpha_2+\beta_2)}{2\pi^2\Gamma(\alpha_1)\Gamma(\alpha_2)\sqrt{x}}\\
&\phantom{=}\hspace{1cm}\times\left[G_{3,3}^{2,3}\left(\left.\begin{array}{c}1,\alpha_2-\beta_1+1/2,\alpha_1-\beta_2+1/2\\\alpha_1-1/2,\alpha_2-1/2,0\end{array}\right|t\right)\right]_{t=x}^{t=1},
}[EqnPDF0ab]
where $G_{p,q}^{m,n}\left(\left.\begin{array}{c}a_1,\ldots,a_p\\ b_1,\cdots,b_q\end{array}\right|z\right)$ is the Meijer G-function.

\subsection{\texorpdfstring{$N=2$}{N=2}}\label{SsRInv2}

Since the odd moments of $j$ vanish, it is convenient to define the variable $x=16j^2$ defined on $[0,1]$.  Using the definition \eqref{Eqnj2} for $j$, the measures \eqref{EqnMeasure2} and the identities \eqref{EqnInt}, it is easy to compute the moments of $x$,
\eqna{
\langle x^n\rangle_\text{Haar}&=\frac{2^{6n}[\Gamma(n+1/2)]^2\Gamma(2n+1)}{\pi\Gamma(4n+2)}=\frac{(1/2)_n^2}{(3/4)_n(5/4)_n},\\
\langle x^n\rangle_\text{Uniform}&=\frac{2^{6n}[\Gamma(n+1/2)\Gamma(2n+1/2)]^2}{\pi^2\Gamma(2n+1)\Gamma(4n+1)}=\frac{(1/4)_n(3/4)_n}{(1)_n^2},
}[EqnM2]
where the duplication formula for the Pochhammer symbol \eqref{EqnDup} was used in the last equalities.  From \eqref{EqnRV}, it is clear that the moments of $x$ \eqref{EqnM2} are the same than the moments of $XY$ for the two random variables $X$ and $Y$ with pdfs $P_{(\alpha_1,\beta_1)}$ and $P_{(\alpha_2,\beta_2)}$ respectively with $(\alpha_1,\beta_1;\alpha_2,\beta_2)=(1/2,1/4;1/2,3/4)$ for the Haar measure and $(\alpha_1,\beta_1;\alpha_2,\beta_2)=(1/4,3/4;3/4,1/4)$ for the uniform measure.  Hence, from \eqref{EqnPDFab}, the pdfs for $x$ are given by
\eqna{
P_\text{Haar}(x)&=P_{(1/2,1/4;1/2,3/4)}(x)=\frac{1}{2\sqrt{2x}}{}_2F_1(1/4,3/4;1;1-x),\\
P_\text{Uniform}(x)&=P_{(1/4,3/4;3/4,1/4)}(x)=\frac{1}{\pi\sqrt{2}x^{1/4}}{}_2F_1(3/4,3/4;1;1-x).
}[EqnPDF2]
Here, Euler's reflexion formula $\Gamma(z)\Gamma(1-z)=\pi/\sin(\pi z)$ was used to simplify the prefactors.

\subsection{\texorpdfstring{$N=3$}{N=3}}\label{SsRInv3}

The logic for the physical case with $N=3$ is essentially the same.  Due to the vanishing of the odd moments of $j_D$, $j_1$ and $j_2$, it is again useful to introduce the variables $x_D=108j_D^2$, $x_1=16j_1^2$ and $x_2=16j_2^2$ defined on $[0,1]$.  Using the definitions \eqref{Eqnj3} for the rephasing invariants, the Haar measure \eqref{EqnMeasure3} and the identities \eqref{EqnInt}, the moments are
\eqna{
\langle x_D^n\rangle_\text{Haar}&=\frac{2^{2n+1}3^{3n}\Gamma(n+1/2)[\Gamma(n+1)]^4}{\sqrt{\pi}\Gamma(2n+2)\Gamma(3n+3)}=\frac{(1)_n^2}{(2n+1)(4/3)_n(5/3)_n},\\
\langle x_{1,2}^n\rangle_\text{Haar}&=\frac{2^{4n+1}\Gamma(n+1/2)[\Gamma(2n+1)]^2}{\sqrt{\pi}\Gamma(n+1)\Gamma(4n+3)}=\frac{(1/2)_n^2}{(2n+1)(3/4)_n(5/4)_n},
}[EqnM3Haar]
where the duplication formula \eqref{EqnDup} was used once more in the last equalities.  As can be seen from \eqref{EqnM3Haar}, the random variables $x_1$ and $x_2$ have the same moments, they therefore have the same pdf.\footnote{This feature is not generic to all measures, as can be seen from the results \eqref{EqnPDF3Uniform} with the uniform measure.}  Using \eqref{EqnRV}, it is straightforward to see that the moments \eqref{EqnM3Haar} are the same than the moments of $XY$ for the two random variables $X$ and $Y$ with pdfs $P_{(\alpha,\beta_1;\alpha,\beta_2)}$ and $P_{(1/2,1)}$ respectively with $(\alpha,\beta_1,\beta_2)=(1,1/3,2/3)$ for $x_D$ and $(\alpha,\beta_1,\beta_2)=(1/2,1/4,3/4)$ for $x_1$ and $x_2$.  Since $\beta_1+\beta_2=1$ in all cases, the pdfs for $x_D$, $x_1$ and $x_2$ in the case of the Haar measure are given by
\eqna{
P_{\text{Haar},D}(x)&=P_{(1,1/3;1,2/3;1/2,1)}(x)=-\frac{1}{6\pi\sqrt{3x}}G_{3,3}^{2,3}\left(\left.\begin{array}{c}1,5/6,7/6\\1/2,1/2,0\end{array}\right|x\right)+\frac{2\pi}{3\sqrt{3x}},\\
P_{\text{Haar},1,2}(x)&=P_{(1/2,1/4;1/2,3/4;1/2,1)}(x)=-\frac{1}{8\pi^2\sqrt{2x}}G_{3,3}^{2,3}\left(\left.\begin{array}{c}1,1/4,3/4\\0,0,0\end{array}\right|x\right)-\frac{\pi}{4\sqrt{x}},
}[EqnPDF3Haar]
from \eqref{EqnPDF0ab}, Euler's reflection formula and the identities
\eqn{G_{3,3}^{2,3}\left(\left.\begin{array}{c}1,5/6,7/6\\1/2,1/2,0\end{array}\right|1\right)=4\pi^2,\qquad\qquad G_{3,3}^{2,3}\left(\left.\begin{array}{c}1,1/4,3/4\\0,0,0\end{array}\right|1\right)=-2\pi^3\sqrt{2},}
which follow from the definition of the Meijer G-function in terms of hypergeometric functions.

The pdfs for the uniform measure \eqref{EqnMeasure3} are not as simple.  Indeed, with the help of \eqref{Eqnj3}, \eqref{EqnMeasure3}, \eqref{EqnInt} and \eqref{EqnDup}, the moments are
\eqna{
\langle x_D^n\rangle_\text{Uniform}&=\frac{2^{2n}3^{3n}[\Gamma(n+1/2)]^6\Gamma(2n+1/2)}{\pi^{7/2}\Gamma(n+1)[\Gamma(2n+1)]^2\Gamma(3n+1)}=\frac{(1/2)_n^4(1/4)_n(3/4)_n}{(1)_n^4(1/3)_n(2/3)_n},\\
\langle x_1^n\rangle_\text{Uniform}&=\frac{2^{4n}\Gamma(n+1/2)[\Gamma(2n+1/2)]^2\Gamma(4n+1/2)}{\pi^2\Gamma(n+1)[\Gamma(4n+1)]^2}=\frac{(1/8)_n(3/8)_n(5/8)_n(7/8)_n}{(1)_n^3(1/2)_n},\\
\langle x_2^n\rangle_\text{Uniform}&=\frac{2^{4n}\Gamma(n+1/2)[\Gamma(2n+1/2)]^3}{\pi^2\Gamma(n+1)\Gamma(2n+1)\Gamma(4n+1)}=\frac{(1/4)_n^2(3/4)_n^2}{(1)_n^3(1/2)_n}.
}[EqnM3Uniform]
Hence, from \eqref{EqnRV} the pdf for $x_D$ is identical to the pdf of the product of three random variables $XYZ$ with respective pdfs $P_{(1/2,1/2;1/2,1/2)}$, $P_{(1/2,1/2;3/4,1/4)}$ and $P_{(1/2,1/6;1/4,1/12)}$ while the pdfs for $x_{1,2}$ are given by the pdfs of the product of two random variables $XY$ with respective pdfs $P_{(1/8,3/8;3/8,5/8)}$, $P_{(5/8,3/8;7/8,1/8)}$ for $x_1$ and $P_{(1/4,1/4;1/4,3/4)}$, $P_{(3/4,1/4;3/4,1/4)}$ for $x_2$.  Therefore, the pdfs for the uniform measure are
\eqna{
P_{\text{Uniform},D}(x)&=\frac{1}{\pi^3\sqrt{3x}}\int_x^1\frac{{}_2F_1(1/2,1/2;1;1-x/t)}{t^{5/4}}\\
&\phantom{=}\hspace{1cm}\times\int_t^1\frac{t'^{1/4}\,{}_2F_1(1/2,1/2;3/4;1-t')\,{}_2F_1(1/6,-1/6;1/4;1-t/t')}{(1-t')^{1/4}(t'-t)^{3/4}}dt'dt,\\
P_{\text{Uniform},1}(x)&=\frac{1}{2\pi^2\sqrt{2}x^{1/8}}\int_x^1\frac{{}_2F_1(3/8,7/8;1;1-t)\,{}_2F_1(3/8,3/8;1/2;1-x/t)}{t\sqrt{t-x}}dt,\\
P_{\text{Uniform},2}(x)&=\frac{1}{2\pi^2x^{1/4}}\int_x^1\frac{{}_2F_1(1/4,3/4;1;1-t)\,{}_2F_1(1/4,1/4;1/2;1-x/t)}{t\sqrt{t-x}}dt.
}[EqnPDF3Uniform]
Unfortunately, the integrals in \eqref{EqnPDF3Uniform} do not seem to lead to elegant analytical solutions.


\section{Discussion}\label{SDis}

This section discusses the implications of the CP-violating rephasing invariant pdfs, compares the different measures and verifies that the analytical rephasing invariant pdfs correspond to the numerical ones.  Considering invariance under $j\to-j$, most results are obtained for $|j|$.  In the physical neutrino case, a comparison with experimental data is undertaken and the implications for CP violation are also discussed.

\subsection{\texorpdfstring{$N=2$}{N=2}}\label{SsDic2}

From \eqref{EqnPDF2}, the pdfs for the rephasing invariant $|j|\in[0,1/4]$ are
\eqna{
P_\text{Haar}(|j|)&=2\sqrt{2}\,{}_2F_1(1/4,3/4;1;1-16|j|^2),\\
P_\text{Uniform}(|j|)&=\frac{8\sqrt{2|j|}}{\pi}{}_2F_1(3/4,3/4;1;1-16|j|^2),
}
and the averages for $|j|$ and $|j|^2$ are $\langle|j|\rangle_\text{Haar}=1/(3\pi)$, $\langle|j|^2\rangle_\text{Haar}=1/60$ and $\langle|j|\rangle_\text{Uniform}=1/(4\pi)$, $\langle|j|^2\rangle_\text{Uniform}=3/256$.  Hence the Haar measure has a larger $|j|$-average than the uniform measure.  Moreover, the $|j|^2$-average for the Haar measure is larger than for the uniform measure, which implies that the standard deviation of the pdf for the Haar measure is larger than for the uniform measure.

The behavior of the pdfs around $|j|=0$ and $|j|=1/4$ is given by
\eqna{
P_\text{Haar}(|j|)&\sim\left\{\begin{array}{ll}-\frac{2}{\pi}\ln(j^2/4)+\cdots&|j|\to0\\2\sqrt{2}-3\sqrt{2}(|j|-1/4)+\cdots&|j|\to1/4\end{array}\right.,\\
P_\text{Uniform}(|j|)&\sim\left\{\begin{array}{ll}\frac{2}{[\Gamma(3/4)]^2}\sqrt{\frac{2}{\pi|j|}}+\cdots&|j|\to0\\\frac{4\sqrt{2}}{\pi}-\frac{10\sqrt{2}}{\pi}(|j|-1/4)+\cdots&|j|\to1/4\end{array}\right.,
}
therefore both pdfs diverge at $|j|\to0$ while they do not vanish at $|j|=1/4$.  For the Haar measure, the divergence at $|j|\to0$ is only logarithmic.

\begin{figure}[!t]
\centering
\resizebox{15cm}{!}{
\includegraphics{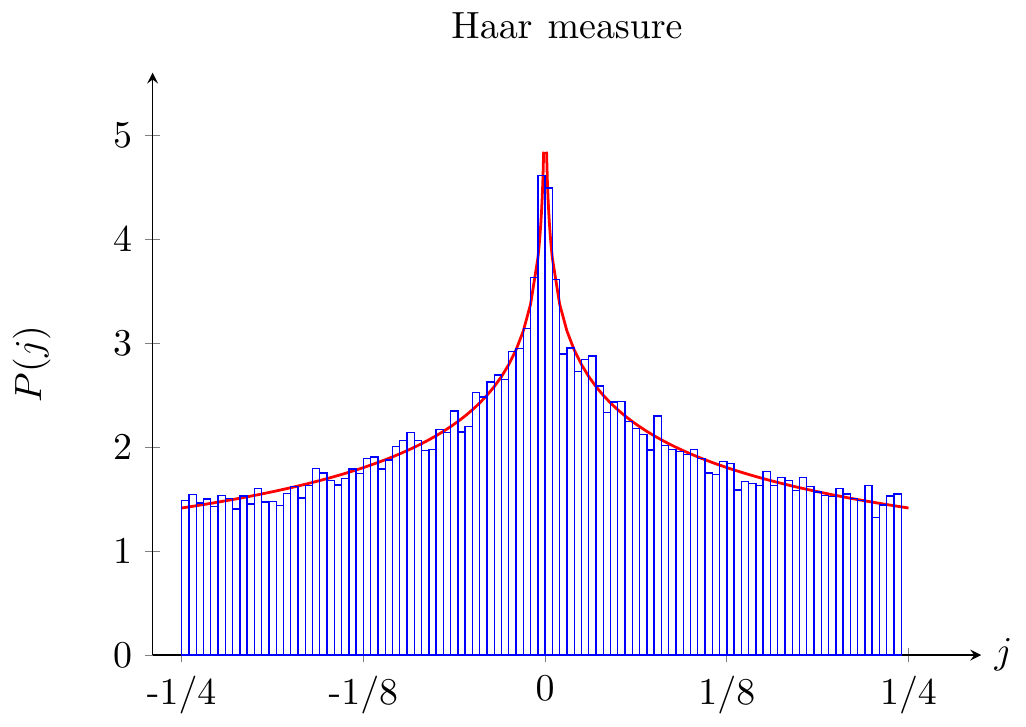}
\hspace{2cm}
\includegraphics{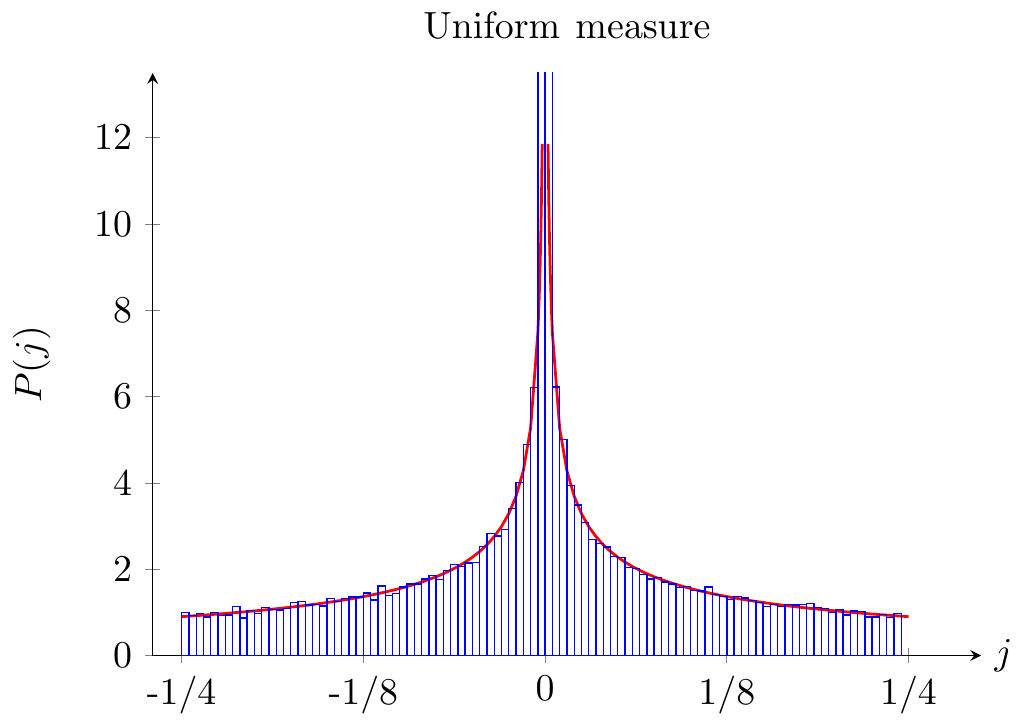}
}
\caption{Probability density functions $P(j)=P(|j|)/2$ of the rephasing invariant $j$ for $2\times2$ unitary matrices.  The red curves correspond to the analytic results while the histograms correspond to numerical results (with $5\times10^4$ unitary matrices).  The left and right panels show the pdfs for the Haar measure and the uniform measure respectively.}
\label{FigPDF2}
\end{figure}

A comparison between the analytic pdfs and numerical results obtained by generating large samples of random unitary matrices is shown in figure~\ref{FigPDF2}.  Figure~\ref{FigPDF2} demonstrates the perfect match between the different results and corroborates the behavior of the pdfs at $|j|=0$ and $|j|=1/4$.

\subsection{\texorpdfstring{$N=3$}{N=3}}\label{SsDic3}

In the physical neutrino case, the three pdfs for the rephasing invariants $|j_D|\in[0,1/(6\sqrt{3})]$, $|j_1|\in[0,1/4]$ and $|j_2|\in[0,1/4]$ are obtained from \eqref{EqnPDF3Haar} for the Haar measure,
\eqna{
P_\text{Haar}(|j_D|)&=-\frac{2}{\pi}G_{3,3}^{2,3}\left(\left.\begin{array}{c}1,5/6,7/6\\1/2,1/2,0\end{array}\right|108|j_D|^2\right)+8\pi,\\
P_\text{Haar}(|j_{1,2}|)&=-\frac{1}{\pi^2\sqrt{2}}G_{3,3}^{2,3}\left(\left.\begin{array}{c}1,1/4,3/4\\0,0,0\end{array}\right|16|j_{1,2}|^2\right)-2\pi,
}
and from \eqref{EqnPDF3Uniform} for the uniform measure,
\eqna{
P_{\text{Uniform}}(|j_D|)&=\frac{12}{\pi^3}\int_{108|j_D|^2}^1\frac{{}_2F_1(1/2,1/2;1;1-108|j_D|^2/t)}{t^{5/4}}\\
&\phantom{=}\hspace{1cm}\times\int_t^1\frac{t'^{1/4}\,{}_2F_1(1/2,1/2;3/4;1-t')\,{}_2F_1(1/6,-1/6;1/4;1-t/t')}{(1-t')^{1/4}(t'-t)^{3/4}}dt'dt,\\
P_{\text{Uniform}}(|j_1|)&=\frac{8|j_1|^{3/4}}{\pi^2}\int_{16|j_1|^2}^1\frac{{}_2F_1(3/8,7/8;1;1-t)\,{}_2F_1(3/8,3/8;1/2;1-16|j_1|^2/t)}{t\sqrt{t-16|j_1|^2}}dt,\\
P_{\text{Uniform}}(|j_2|)&=\frac{8\sqrt{|j_2|}}{\pi^2}\int_{16|j_2|^2}^1\frac{{}_2F_1(1/4,3/4;1;1-t)\,{}_2F_1(1/4,1/4;1/2;1-16|j_2|^2/t)}{t\sqrt{t-16|j_2|^2}}dt.
}
In order to compare the pdfs for the two measures, the lowest moments for $|j_D|$, $|j_1|$ and $|j_2|$ are shown in table~\ref{TabMoments}.
\begin{table}[!t]
\centering
\resizebox{7cm}{!}{%
\begin{tabular}{|c||c|c|}
\hline
 & Haar measure & Uniform measure\\\hline
$\langle|j_D|\rangle$ & $\pi/105$ & $4/(3\pi^4)$\\
$\langle|j_1|\rangle$ & $1/(6\pi)$ & $3/(32\pi)$\\
$\langle|j_2|\rangle$ & $1/(6\pi)$ & $1/(8\pi)$\\
$\langle|j_D|^2\rangle$ & $1/720$ & $1/2048$\\
$\langle|j_1|^2\rangle$ & $1/180$ & $105/32768$\\
$\langle|j_2|^2\rangle$ & $1/180$ & $9/2048$\\
\hline
\end{tabular}
}
\caption{Lowest moments for the rephasing invariants $|j_D|$, $|j_1|$ and $|j_2|$ for the Haar measure and the uniform measure.}
\label{TabMoments}
\end{table}
Table~\ref{TabMoments} shows that the average values for the absolute values of the CP-violating rephasing invariants are always larger for the Haar measure than for the uniform measure.  Moreover, it is clear from table~\ref{TabMoments} that the pdf standard deviations for the Haar measure are larger than the ones for the uniform measure.

\begin{figure}[!t]
\centering
\resizebox{15cm}{!}{
\includegraphics{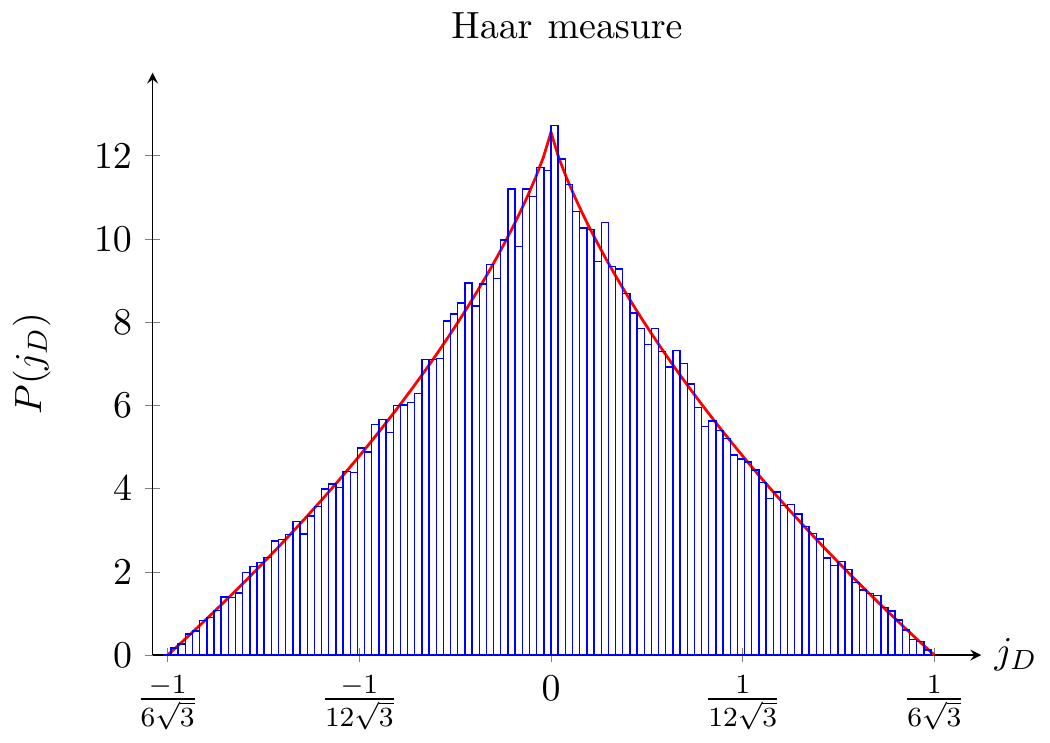}
\hspace{2cm}
\includegraphics{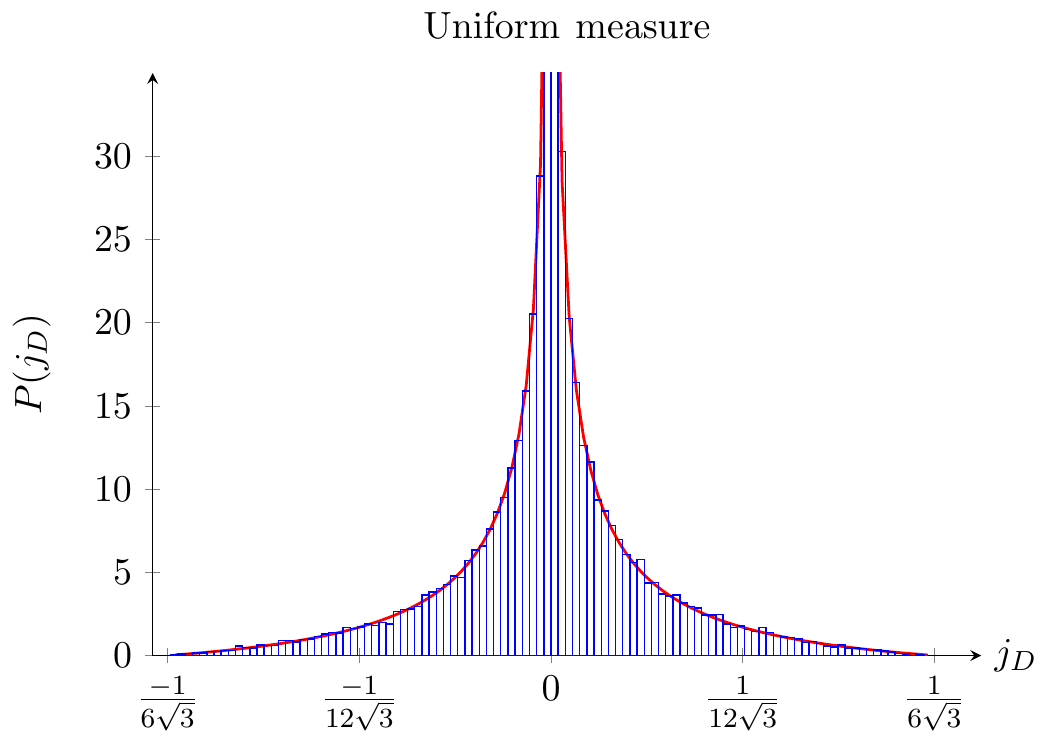}
}
\resizebox{15cm}{!}{
\includegraphics{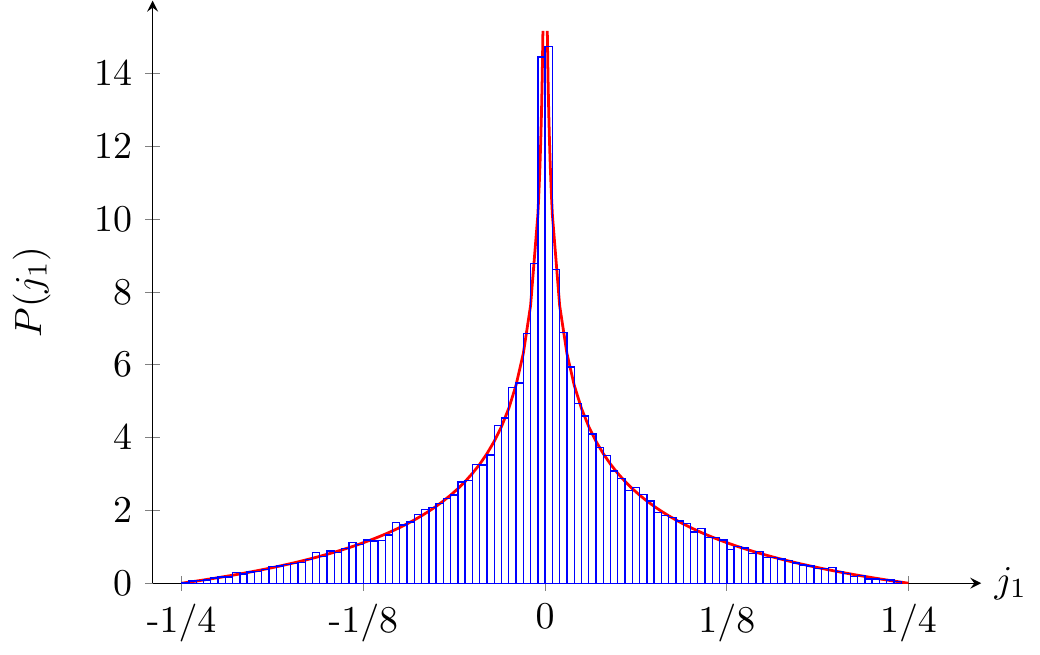}
\hspace{2cm}
\includegraphics{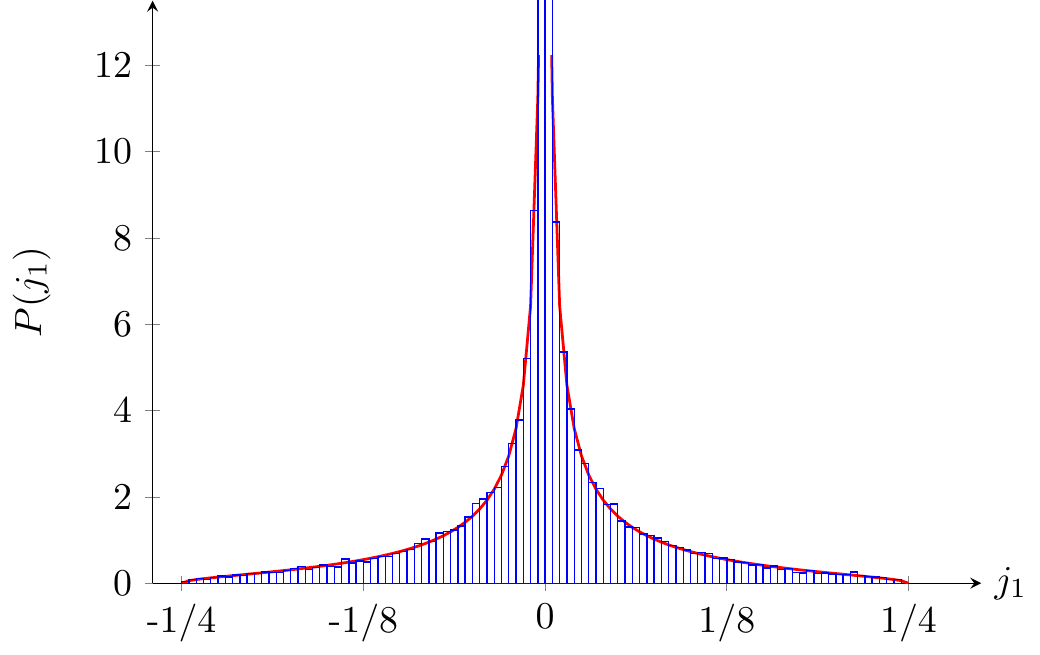}
}
\resizebox{15cm}{!}{
\includegraphics{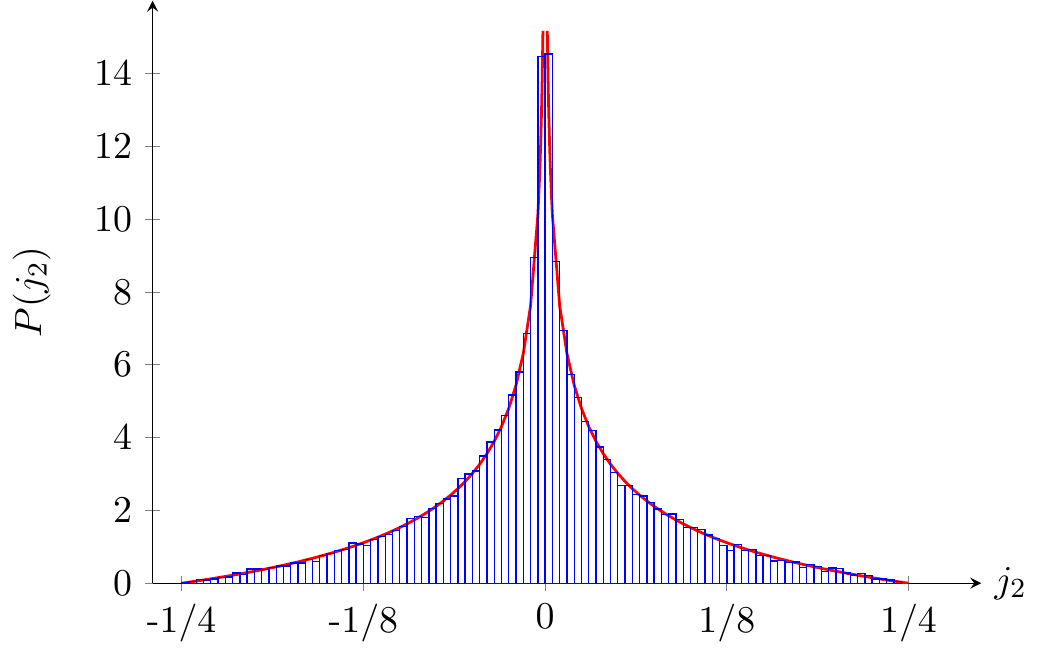}
\hspace{2cm}
\includegraphics{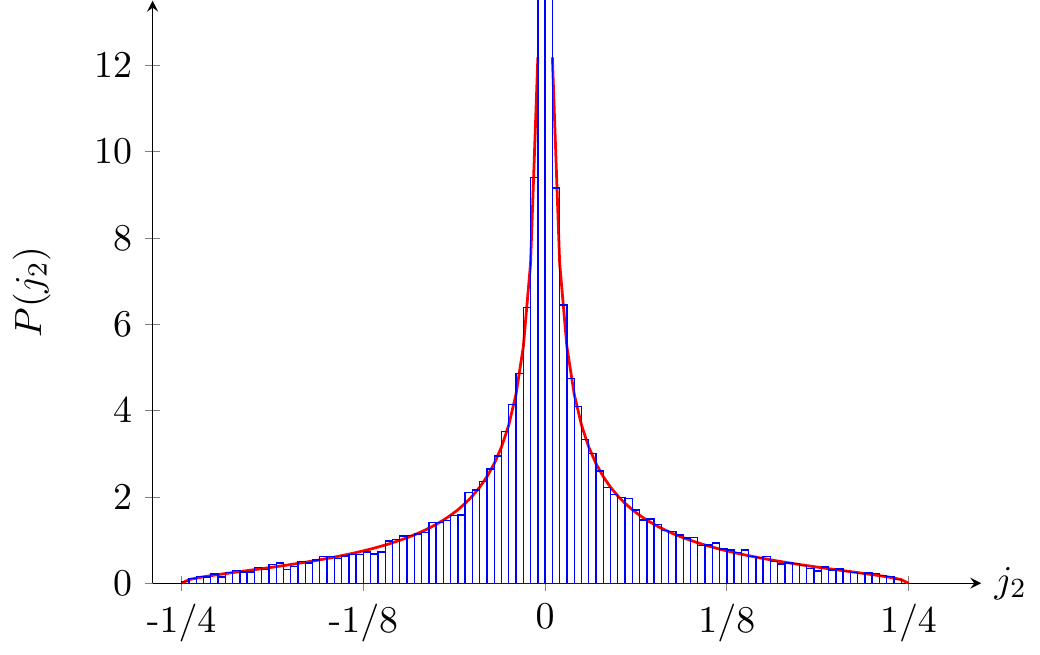}
}
\caption{Probability density functions $P(j)=P(|j|)/2$ of the rephasing invariants $j_D$ (top panels), $j_1$ (middle panels) and $j_2$ (bottom panels) for $3\times3$ unitary matrices.  The red curves correspond to the analytic results while the histograms correspond to numerical results (with $5\times10^4$ unitary matrices).  The left and right panels show the pdfs for the Haar measure and the uniform measure respectively.}
\label{FigPDF3}
\end{figure}
Expanding around the endpoints, it is easy to verify that the pdfs vanish at the endpoints for both the Haar measure and the uniform measure, contrary to the case $N=2$.  Moreover, the CP-violating Dirac phase pdf for the Haar measure with $N=3$ reaches a maximum value at the origin of $8\pi$.  As for the $N=2$ case, the remaining pdfs diverge around the origin.

A comparison between the analytical results and the numerical results, again obtained with the help of large samples of random unitary matrices, is performed in figure~\ref{FigPDF3}.  The agreement of the analytical results with the samples of randomly-generated unitary matrices demonstrates the validity of the method.

Before concluding, it is interesting to analyse the implications of the pdfs computed above in the physical neutrino case.  First, since the pdfs cannot determine the sign of the CP-violating rephasing invariants, the analysis performed here will focus on their absolute values.  Hence, the moments of interest are the ones given in table~\ref{TabMoments}.
\begin{table}[!t]
\centering
\resizebox{8cm}{!}{%
\begin{tabular}{|c||c|c|}
\hline
 & Normal hierarchy & Inverted hierarchy\\\hline
$|j_D^\text{max}|$ & $0.0329_{-0.0007}^{+0.0007}$ & $0.0328_{-0.0007}^{+0.0007}$\\
$|j_1^\text{max}|$ & $0.203_{-0.004}^{+0.005}$ & $0.203_{-0.004}^{+0.005}$\\
$|j_2^\text{max}|$ & $0.0147_{-0.0006}^{+0.0006}$ & $0.0148_{-0.0006}^{+0.0006}$\\
\hline
\end{tabular}
}
\caption{Best-fit maximum values for the CP-violating rephasing invariants of the PMNS neutrino matrix for the normal and inverted hierarchies \cite{Esteban:2016qun}.  The intervals correspond to $\pm1\sigma$.}
\label{TabNeutrinos}
\end{table}

Second, from the experimental values for the maximum values of the CP-violating rephasing invariants shown in table~\ref{TabNeutrinos}, it is possible to compute the probability that a given measure generates an allowed rephasing invariant by integrating over the permitted range $|j_D|\in[0,|j_D^\text{max}|]$, $|j_1|\in[0,|j_1^\text{max}|]$ or $|j_2|\in[0,|j_2^\text{max}|]$.  This corresponds to allowing all possible values between $0$ and $2\pi$ for the CP-violating phases.  These probabilities $P\{|j|\leq|j^\text{max}|\}$, which are integrated pdfs up to $|j|=|j^\text{max}|$, are given in table~\ref{TabProb}.
\begin{table}[!t]
\centering
\resizebox{15cm}{!}{%
\begin{tabular}{|c||cc|cc|}
\hline
 & \multicolumn{2}{|c|}{Haar measure} & \multicolumn{2}{|c|}{Uniform measure}\\
 & Normal hierarchy & Inverted hierarchy & Normal hierarchy & Inverted hierarchy\\\hline
$P\{|j_D|\leq|j_D^\text{max}|\}\,(\%)$ & $60.7\pm0.9$ & $60.6\pm0.9$ & $86.5\pm0.4$ & $86.4\pm0.4$\\
$P\{|j_1|\leq|j_1^\text{max}|\}\,(\%)$ & $98.6\pm0.3$ & $98.6\pm0.3$ & $98.8\pm0.2$ & $98.8\pm0.2$\\
$P\{|j_2|\leq|j_2^\text{max}|\}\,(\%)$ & $29.8\pm0.8$ & $29.9\pm0.8$ & $49.4\pm0.7$ & $49.5\pm0.7$\\
\hline
\end{tabular}
}
\caption{Probabilities $P\{|j|\leq|j^\text{max}|\}$ for the Haar measure and the uniform measure that the CP-violating rephasing invariants are in the allowed experimental ranges for the normal and inverted hierarchies.}
\label{TabProb}
\end{table}
Since all the probabilities are quite large, they demonstrate that the statistical hypothesis that the PMNS matrix arises randomly from a probability experiment with the Haar measure or the uniform measure cannot be rejected.  Moreover, a comparison between the probabilities for the Haar measure and the uniform measure shows that the uniform measure is slightly preferred over the Haar measure, although a statistical test like the likelihood-ratio test would not give significant statistical evidence to discriminate between the two measures.

Finally, a comparison of the averages of the absolute values of the rephasing invariants shown in table~\ref{TabMoments} and the experimental values of table~\ref{TabNeutrinos} leads to predictions for the CP-violating phases $\delta$, $\alpha_{21}$ and $\alpha_{31}$.  These predictions are shown in table~\ref{TabPhases}.
\begin{table}[!t]
\centering
\resizebox{14cm}{!}{%
\begin{tabular}{|c||cc|cc|}
\hline
 & \multicolumn{2}{|c|}{Haar measure} & \multicolumn{2}{|c|}{Uniform measure}\\
 & Normal hierarchy & Inverted hierarchy & Normal hierarchy & Inverted hierarchy\\\hline
$\delta\,({}^\circ)$ & $65.4_{-2.7}^{+2.7}$ & $65.8_{-2.7}^{+2.7}$ & $24.7_{-0.6}^{+0.6}$ & $24.7_{-0.6}^{+0.6}$\\
$\alpha_{21}\,({}^\circ)$ & $15.1_{-0.3}^{+0.4}$ & $15.1_{-0.3}^{+0.4}$ & $8.45_{-0.17}^{+0.21}$ & $8.45_{-0.17}^{+0.21}$\\
$\alpha_{31}\,({}^\circ)$ & $-$ & $-$ & $-$ & $-$\\
\hline
\end{tabular}
}
\caption{Predicted first quadrant values of the CP-violating phases from the pdfs with the Haar measure and the uniform measure for the normal and inverted hierarchies.  It is important to note that since only the absolute values of the sine of the phases are constrained, there are three additional values allowed for each phases, given by $180^\circ\pm90^\circ\pm(90^\circ-\phi)$ where $\phi\in\{\delta,\alpha_{21},\alpha_{31}\}$.}
\label{TabPhases}
\end{table}
Considering that the experimental hint for the Dirac CP-violating phase $\delta$ is $(261_{-59}^{+51})^\circ$ for the normal hierarchy or $(277_{-46}^{+40})^\circ$ for the inverted hierarchy \cite{Esteban:2016qun}, it is clear that the Haar measure (and also the uniform measure) is compatible with the experimental value $|j_D^\text{exp}|$ for both hierarchies when considering the two appropriate possible values in table~\ref{TabPhases} for $\delta$.  The predictions for the Majorana CP-violating phases lead to a small value for the sine of the phase $\alpha_{21}$ and no acceptable answer for $\alpha_{31}$.  The latter behavior originates from the average value of $|j_2|$ found in table~\ref{TabMoments} being larger than the maximum experimental value of table~\ref{TabNeutrinos}.  Taking into account $1\sigma$ variations on the average value of $|j_2|$ found in table~\ref{TabMoments}, all possible values of the phase $\alpha_{31}$ are then allowed.  Therefore, no precise prediction on $\alpha_{31}$ can be made from the analysis presented here.

With respect to CP violation, the predictions obtained here indicate that the central experimental value for the CP-violating Dirac phase $\delta$ is likely correct while the CP-violating Majorana phase $\alpha_{21}$ is small (although the associated rephasing invariant $|j_1|$ is relatively large).  Therefore, CP violation in the lepton sector would mostly originate from the CP-violating Dirac phase $\delta$ unless the CP-violating Majorana phase $\alpha_{31}$, which is unconstrained here, ends up being quite large (although the associated rephasing invariant $|j_2|$ is relatively small).  Furthermore, the fact that the prediction for the CP-violating Majorana phase $\alpha_{21}$ is not a multiple of $\pi$ indicates that $\alpha_{21}$ breaks CP, with obvious implications for neutrinoless double $\beta$-decay.

From the agreement for $|j_D|$, it is therefore possible to argue that the PMNS matrix is a generic unitary matrix obtained randomly from a statistical ensemble generated by the Haar measure.  This observation is contrary to the CKM case \cite{Dunkl:2009sn}.  Indeed, for the CKM matrix, the observed CP-violating Dirac phase is $|j_D^\text{exp}|_\text{CKM}=(3.04_{-0.20}^{+0.21})\times10^{-5}$ \cite{pdg} with the very small probability $P\{|j_D|\leq|j_D^\text{exp}|_\text{CKM}\}=(7.64_{-0.51}^{+0.52})\times10^{-4}\approx0.08\%$ for the Haar measure.  Hence, as opposed to the quark sector of the SM that is highly hierarchical, the lepton sector of the SM could very well have its origin in the anarchy principle that is the original physical motivation for the Haar measure presented here.  Moreover, CP violation would be more important in the lepton sector, suggesting for example that the baryon asymmetry of the Universe is indeed due to leptogenesis.


\section{Conclusion}\label{SConclusion}

This paper investigated the implications of the anarchy principle for CP violation in the lepton sector through the CP-violating rephasing invariants.  After a short introduction of the motivations for the anarchy principle in the lepton sector and a quick discussion of the rephasing invariants, the pdfs for the CP-violating rephasing invariants were obtained.  These pdfs were computed with the help of their moments for both the Haar measure relevant to the anarchy principle and the uniform measure.  Using some elementary mathematical results, the appropriate pdfs were obtained as the pdfs of products of random variables.  Elegant analytical results were given for both measures in the $N=2$ case and for the Haar measure in the $N=3$ case.\footnote{It is obvious that straightforward integrations of the corresponding measure with appropriate changes of variables lead to correct pdfs.  However, this brute-force technique cannot be easily generalized.}  A discussion of the physical implications of the Haar measure for neutrino physics was also presented.

The most important observation that was made is that the PMNS matrix, contrary to the CKM matrix, is very likely to be a generic unitary matrix generated randomly by the statistical ensemble associated to the Haar measure.  Indeed, for the normal hierarchy, a comparison of the observed leptonic (quarkonic) Jarlskog invariant $|j_D^\text{exp}|=0.032_{-0.005}^{+0.005}$ ($|j_D^\text{exp}|_\text{CKM}=(3.04_{-0.20}^{+0.21})\times10^{-5}$) with the average $\langle|j_D|\rangle_\text{Haar}=\pi/105\approx0.030$ shows a striking (dis)agreement between the two.\footnote{The agreement is not as good for the uniform measure where $\langle|j_D|\rangle_\text{Uniform}=4/(3\pi^4)\approx0.014$.}  Moreover, although it was not possible to discriminate between the different measures nor the different hierarchies, the results presented here do suggest that the anarchy principle, or any other mechanism responsible for a PMNS pdf with the Haar measure, could be implemented by Nature.  With that in mind, it will be interesting to see how the prediction on the CP-violating Majorana phase $\alpha_{21}$ presented here holds on with proposed and on-going dedicated experiments probing the neutrino sector.

Finally, it is worth noting that the analysis presented here can be generalized to any $N$.  Indeed, since the CP-violating rephasing invariants are known for all $N$ and a suitable parametrization of the unitary matrix exists (as in \cite{1751-8121-43-38-385306,:/content/aip/journal/jmp/53/1/10.1063/1.3672064}), the moments can in principle be calculated straightforwardly.  From the mathematical results presented here, these moments should be expressible as products of beta-distributed random variables, although there is no guarantee that the final pdfs are simple.  This generalization is of obvious interest for sterile neutrinos and it could also be relevant for leptogenesis, which usually relies on high-energy information about CP violation.


\ack{
This work is supported by NSERC.
}


\bibliography{DistInv}

\end{document}